\documentclass[12pt]{article}
\usepackage{graphicx,epstopdf,amsmath,amssymb,amsthm,euscript,endnotes,color, tipa, dsfont}
\usepackage[russian, english]{babel}

\usepackage{txfonts}
\newcounter{aaa}
\newenvironment{teor}[2][{}]{\begin{trivlist}\refstepcounter{aaa}%
\item[\bfseries \theaaa. #2. ]#1}%
{\end{trivlist}}

\renewcommand{\leq}{\leqslant}
\newcommand*{\obozn}{
\equiv }

\newcommand{\ssy}[6]{#1, {\it #2}\ #3 {\bf #4} (#5) #6\rlap{.}}

\newcommand{\rmd}{\mathrm{d}}
\DeclareMathAlphabet{\bi}{OML}{cmm}{b}{it}
\DeclareSymbolFont{iso}{U}{txmia}{m}{it}
\DeclareMathSymbol{\cel}{\mathalpha}{iso}{"9A}
\DeclareMathSymbol{\mink}{\mathalpha}{iso}{"8C}
\DeclareMathSymbol{\eucl}{\mathalpha}{iso}{"85}
\DeclareMathSymbol{\rea}{\mathalpha}{iso}{"92}
\DeclareMathSymbol{\nat}{\mathalpha}{iso}{"8E}

\DeclareMathOperator{\Cl }{Cl}%
\begin{document}

\title{A causality preserving evolution of a  pair of strings}
\author{S. Krasnikov\thanks{krasnikov.xxi@gmail.com}\\
Central Astronomical Observatory at Pulkovo, St.~Petersburg,\\ 196140,
 Russia}
\maketitle
\begin{abstract}
As  follows from Gott's discovery, a pair
of straight string-like singularities moving in opposite directions, when they have  suitable  speed and impact parameter, produce
  closed timelike curves.
  I argue in this paper that there always is a not-so-frightening alternative: the Universe may prefer to produce a certain (surprisingly simple and absolutely mild) singularity instead.
\end{abstract}

\begin{flushright}
    \emph{To Milena}
\end{flushright}
\section{Introduction}

\subsection{The objective}
The subject matter of this paper is some properties of a very special type of singularity. Suppose,  a spacetime
admits a finite open  convex  covering
\begin{equation}\label{eq:cov}
M=\bigcup_{i=1,\dots i_0<\infty} M_i
\end{equation}  such that each  $ M_i  $ is a   subset of a Minkowski $n$-dimensional  space $\mink^n$. Such spacetimes are exceptionally simple and this enables one  \cite{strstr} to assign a ``form" to singularities contained  there
(throughout the paper we stick to the physical level of rigor and drop discussion of ``self-evident" concepts and facts).
When such a singularity has the form of an $(n-2)$-dimensional surface, the former is called a \emph{string-like singularity} or just  a \emph{string}.

String-like singularities
are abundant in GR, see \cite{mee,SheVi,Braz} for some reviews and references, and    \cite{strstr} for singularities \eqref{eq:cov} of less trivial forms.
For instance, in the $(n=3)$ case the singularities are associated with world lines of massive pointlike particles and
  in the $(n=4)$ case their ``surface-like" counterparts--\emph{approximately}--describe cosmic strings (for an example of a use of this approximation see \cite{Gott}).
\\
\textbf{Warning.}  Still, string-like singularities are not to be confused with cosmic strings. They are objects of different  nature. The former, in particular, are purely geometric entities. In contrast to the latter  they are 
 \emph{everywhere} flat, which implies, in particular, that they solve the vacuum Einstein  equations.   Also, they do not bend or, say, emit gravitational waves.\\

Not much is known about these strings' possible  dynamics. Until recently the only relevant  result was that obtained by Hellaby \cite{Hell} who proved that mutually perpendicular string-like singularities  do not pass intact  through each other.
Another, almost concurrent,   result was due to Gott \cite{Gott}. Loosely speaking,  it says
that two parallel  strings in  $\mink^4$ or, equivalently, 2D cones in  $\mink^3$ having the  angle deficit $\alpha $ and
moving in opposite directions with some  particular speed $v<c$,  and impact parameter $d$  (a \emph{Gott pair}), produce
  closed timelike curves. True, it was claimed in \cite{djh} that the initial conditions at spacelike infinity
  of Gott's spacetime are unphysical, but as pointed out by  Headrick and Gott \cite{hg}   unphysicality in \cite{djh}
    is postulated rather than derived,
The goal of this paper is to prove by construction that the causality violation is unnecessary: for any $d$,  $\alpha $, and $v$ there is an inextendible  
  spacetime, $S_f$, which describes the  said scattering, but in which the causality condition holds everywhere.

\subsection{Gluing spacetimes}

All relevant spacetimes below are built from  Minkowski space by a certain
 manipulation called ``gluing" and in this subsection we  provide a tolerably formal meaning to that word  in application to
portions of spacetimes.

\begin{figure}[th]
\begin{center}
  \includegraphics[width=\textwidth]{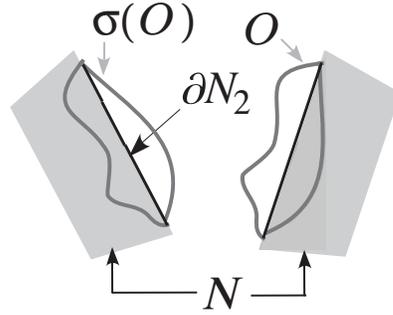}
\end{center}
  \caption{\label{fig:skl3}
The white area is $U$.  }
\end{figure}

From a 
 spacetime $M$ remove   a subset $U$ such that
the space $N\obozn M-U$ is a spacetime with boundary. Now suppose that for some    components of the boundary,
$\partial N_1$, $\partial N_2$ there is  a neigh\-bour\-hood $O\supset\partial N_1$ and  an 
    embedding
$\sigmaup\colon O\to M$  such that
\begin{equation}\label{eq:glu}
O \cap \sigmaup(O) = \varnothing,\qquad
 \sigmaup (O- N)\subset N, \qquad
 \sigmaup(\partial N_1 ) = \partial N_2.
\end{equation}
Then   \begin{equation}\label{eq:S}
S\equiv   N \cup _ {\sigmaup} O
\end{equation}
is said to be the result of  gluing together $\partial N_1$ and $\partial N_2$, cf.~figure~\ref{fig:skl3}, by $\sigmaup$.

\section{Elementary examples}
Now   we are going to consider a couple of simple specific examples. They are interesting by themselves and also they will be useful in the next section.

\begin{teor}{Example. Straight  string}\label{ex:glu}
Let $U$  be the dihedral  angle 
  swept in the course of evolution by a resting 2D cone   $\mathcal{A}$
\begin{equation}\label{eq: 2D}
U=\bigcup_t \mathcal{A}_{t},\quad   \mathcal{A}_{t_0}\obozn \{p:\qquad t(p)=t_0, \quad - \alpha^* /2 < \phi(p)<\alpha^* /2 \}.
\end{equation}
in Minkowski space
  \begin{equation}\label{eq:cone metr}
\rmd s^2= -\rmd t^2 +  \rmd \ r^2 + \ r^2\rmd \phi^2,\qquad
  t \in \rea,\quad\ r>0,\quad\phi \text{ is identified with }\phi+2\pi
\end{equation}
where   $\alpha^*  $  is a non-zero constant smaller than $\pi$.
\begin{figure}[h!tb]
  \includegraphics[width=0.8\textwidth]{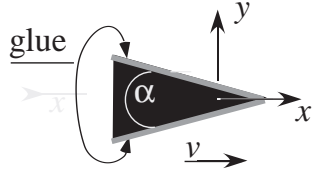}
  \caption{The black angle is the ``hole" $\mathcal{B}_{t_0}$ for a certain $t_0$ (to convert  the  former to   $\mathcal{A}_{t_0}$ set $v=0$, $\alpha=\alpha^*$). $S_B$ is obtained by identifying---at each $t$---the upper gray ray, $y=-\tg(\alpha/2)(x-vt), \  y>0$,   with the lower one,  $y=\tg(\alpha/2)(x-vt), \  y<0$. }
 \label{fig:kart_12}
\end{figure}
Then the    components of the boundary are the half planes
$\partial N_{1,2} =  \{p:\ \phi(p) = \mp \alpha^* /2 \}$,
the neighbourhood  $   O$ can be chosen to be the wedge $ \{p:\ - \alpha^* /2 - \epsilon < \phi(p)<- \alpha^* /2 +\epsilon \}$ and $\sigmaup$ be the rotation by $\alpha^*  $ in the planes of fixed $t$. It is the thus obtained spacetime $S$---referred to as $S_A$ in this particular case---that is usually called string.

\end{teor}

\begin{teor}{Example.  Running angle. Fig~\protect{\ref{fig:kart_12}}}\label{ex:run}

%
%
%
%



In order to construct  a mild generalization   (depicted in Fig~\ref{fig:kart_12})  of  the string singularity discussed above, let us redefine $U$
\begin{equation}\label{eq: 2D1}
U\obozn\bigcup_t \mathcal{B}_{t},\qquad    \mathcal{B}_{t_0}\obozn \ \tauup_{t_0}(\mathcal{A}_{t_0}),
\end{equation}
where $\tauup(\cdot)_{t_0}$ is the translation  by $v{t_0}$ in the $x$-direction ($x$ being a cartesian coordinate $x\obozn r\cos \phi$) with $v\obozn const<c$ and $\bi v \sim \vec{Ox}$. Obviously, the just built spacetime, $S_B$, 
and the former one, $S_A$, are \emph{isometric}, they are related by a boost 
in the $x$-direction.

Two  circumstances are especially noteworthy:
 \begin{enumerate}
   \item pick a point $p$ of the upper gray ray in Fig.~\ref{fig:kart_12}. The deficit angle 
   of the moving  string, $\alpha$, is
   \begin{equation*}
    \alpha = 2\arctan    [  y(p)/ x(p)]
    \end{equation*}
 whence
 \begin{equation*}
    \alpha ^*= 2\arctan    [  y^*(p)/ x^*(p)]
    = 2\arctan   \{   y(p)/[ \gamma(v)x(p)] \}.
    \end{equation*}
The asterisk here denotes "in the proper  reference system  of the string" and.$\gamma$ is the Lorentz factor.  Thus
\begin{equation}\label{eq:bar a}
\tg \frac{\alpha}2=\frac 1{\sqrt{1-v^2}}   \tg \frac{\alpha^*} 2;
\end{equation} 
  \item
 the rotation axis of the moving string $S_B$ is not parallel to the $t$-axis. Correspondingly, a vector initially lying in the $(x,y)$-plane acquires, after being transported around the string,   a non-zero $t$-component. This means, in particular, that $t$ is a ``bad" coordinate: the  (maximal extensions of) surfaces $t=const$ are not embedded into the spacetime.
\end{enumerate}
%

\end{teor}

\begin{teor}{Example. Inelastic head-on collision}

Now consider  the spacetime with \label{sec:inel}
\begin{equation}\label{eq: 1+1}
U=U_C \obozn \bigcup_t \mathcal{C}_{t},\qquad    \mathcal{C}_{t_0}\obozn \ \mathcal{B}_{-t_0} \cup\rhoup_Y(\mathcal{B}_{-t_0}),
\end{equation}
where $ \mathcal{B}$ are defined in \eqref{eq: 2D1} and  $\rhoup_Y(\cdot)$ is the    reflection  through the $y$-axis    in  the $(x,y)$-plane. $U$  at each moment of $t$ is a pair of  equal angles moving---with the speed $v$---towards each other until at $t=0$ their vertices  collide, see figure~\ref{fig:lob2}.
At positive $t$,  $\mathcal{B}_{-t}$  and $ \rhoup_Y(\mathcal{B}_{-t})$ start to overlap, see figure~\ref{fig:lob2}c. Or they  can be viewed as a pair of receding obtuse angles (bounded by the gray lines in figure~\ref{fig:lob2}) either  of which has magnitude $\pi - \alpha$ and (vertically directed)  velocity $v\tg \frac\alpha2$.
\begin{figure}[h!tb]
  \includegraphics[width=\textwidth]{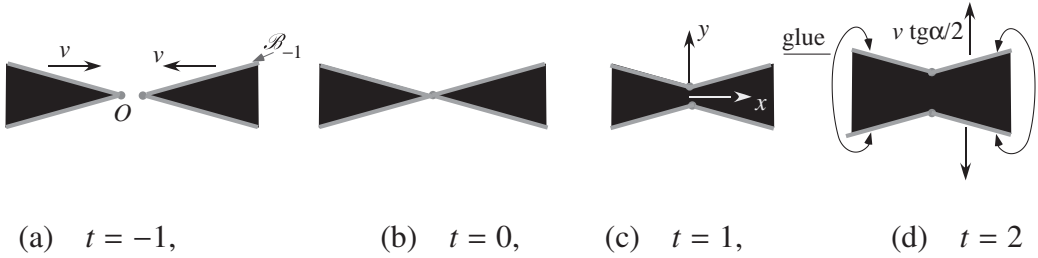}
 \begin{center}
(a)\quad  $t=-1$,\hfil\qquad\qquad \qquad (b)\quad   $t=0$, \qquad\hfil (c)\quad   $t=1$, \hfil\qquad \qquad (d)\quad   $t=2$\hfil
 \end{center}
  \caption{The cross sections $t=const$ of $N$ (the white area) and of $\partial N$ (the gray rays). $S_C$ is a pair of strings merging into a single resting one.}\label{fig:lob2}
\end{figure}
The spacetime $S$, denoted $S_C$ in this case, describing the head-on collision of two cones (or two parallel strings)
 is obtained by the pairwise gluing together---at each $t$---the upper   two gray rays $y=\tg(\alpha/2)(|x|+vt), \  y>0$  with the lower  
two     $y=-\tg(\alpha/2)(|x|+vt), \  y<0$.

$S_C$  contains no closed causal curves. This observation  is not quite trivial as is seen from the comparison between $S_C$ and a Gott pair. The proof is based on the fact that $\sigmaup$
obeys the condition
\begin{equation}\label{eq:iso}
    t(\sigmaup(p))=t(p) ,  \qquad \forall p\in \partial N_i
    \end{equation}
 (i.~e., only points with the same $t$ are identified). $t$ grows along any future directed causal curve and hence   such a curve cannot be closed.

\end{teor}

\section{String---string scattering}

In this section we finally present a spacetime  that can be interpreted as a causality respecting evolution of a Gott pair.

\begin{figure} 
\begin{center}
 \includegraphics[width=\textwidth]{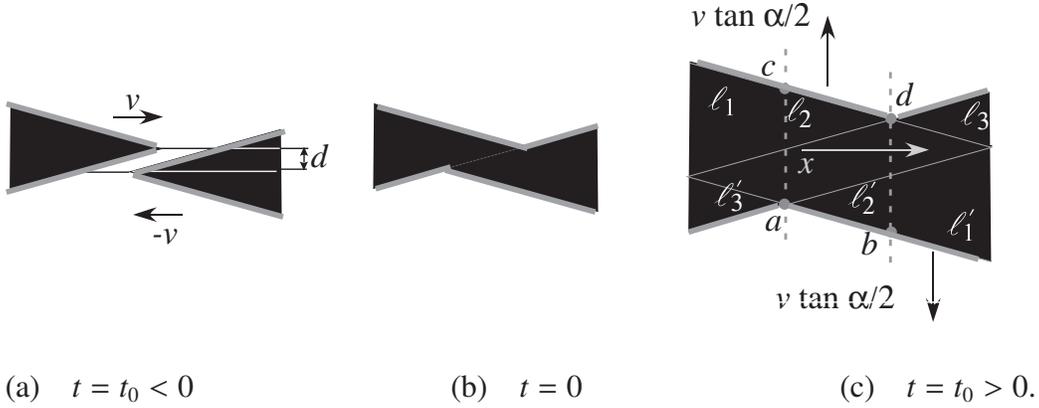}%
\end{center}%
 (a)\quad  $t=t_0<0$\hfill (b)\quad   $t=0$\hfill (c)\quad  $t=t_0>0$.\qquad
 \caption{The gray lines are the sides of corresponding angles.}\label{fig:evol}
\end{figure}

We start with  the spacetime $U_X\subset  \mink^3$ which
differs from $U_C$, see~eq.~\eqref{eq: 2D1}, by one detail: the
   translation  $\tauup $ is changed to the superposition $ \deltaup \circ \tauup$, where $\deltaup$ is the translation  by $d/2$ in the $y$-direction. The cross-section $t=t_0<0 $  of $U_X$ 
    is a pair of angles, moving towards
 each other  with speed $v$ and
   with non-zero
 impact parameter $d$, see figure~\ref{fig:evol}a.
 At positive $t$'s  the angles partially overlap taking  the form of a skewed bowtie, see figure~\ref{fig:evol}c. The bowtie's boundary is a pair of broken lines, related by the point reflection $\omegaup$ through the origin, of which  the upper one   consists of the straight  segments $\ell_1\obozn(-\infty,c),\ \ell_2\obozn(c,d),\ \text{and } \ell_3\obozn(d,\infty)$. Correspondingly, the lower
 broken line is constituted by the segments $\ell'_i\obozn\omegaup(\ell_i)$, $i=1..3$.
 In the course of evolution, all four vertices $a,b,c,d$ change  their location and  each $\ell_i^{(\prime)}$ sweeps a strip  $\EuScript L_i^{(\prime)}$.

\begin{figure}
  \includegraphics[width=\textwidth]{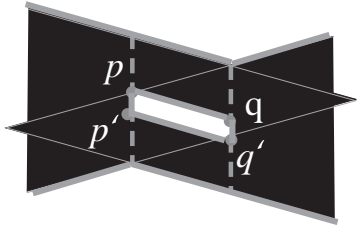}\\
  \caption{The cross section of the middle part of $S_f$. At each positive $t$ the following segments are identified: $(p'q')=l'_2$, $(pq)=l_2$,  $(p'p)=(q'q)$.
  }\label{fig:evol2}
\end{figure}

 Now glue $\EuScript L_3^\prime$  to $\EuScript L_1$ and
 $\EuScript L_3$  to $\EuScript L_1^{\prime}$ (the gluing isometries being the rotation with the duly tilted axes, cf.~example~\ref{ex:run}). The resulting spacetime, $R$,
 has almost all properties, cf.~Conclusions, of the sought-for spacetime  $S_f$. The former, however, is extendible (that is  there exists a spacetime $X$ ``greater" than $R$, i.~e., $R \varsubsetneq X$). To eliminate this last ``flaw",
  let us, first, introduce one more object---the parallelogram $(p',p,q,q')$
  depicted by
  the white quadrangle in figure~\ref{fig:evol2}
     and defined as
   the  parallelogram $(a,c,d,b)$,  see figure~\ref{fig:evol}c, contracted in the vertical direction so that
\begin{equation}\label{eqe^zazor}
 \begin{split}
 \forall t&\leq 0\qquad (p',p,q,q') 
 =\varnothing,
 \\
\forall t&>0\qquad y(c)-y(p)=y(p')-y(a)=const.
\end{split}
\end{equation}
Of course the locations of the points $p^{(\prime)}$, $q^{(\prime)}$ are again functions of $t$. So, with the passage of time the segments
\[
 (p'q'), \quad    (p'p), \quad      (qq')  , \quad  (pq)
 \]
 sweep four strips which we shall denote  by
$\EuScript K$ with   
corresponding indexes.

The last step in building $S_f$ is gluing together
\[\EuScript K _{pq}  \text{ and }\EuScript L_{2}, \quad
\EuScript K _{p'q'}  \text{ and }\EuScript L'_{2}, \quad
\EuScript K _{pp'}  \text{ and } K _{qq'}
\]
which is possible because, as follows  directly from 
the definition,  $\EuScript K _{pq}  $,
$\EuScript K _{p'q'}$, and $\EuScript K _{p{p^{\prime}  } } $  have the  same  proper widths   as and are
parallel to $\EuScript L_2$, $\EuScript L_{2}'$, and $\EuScript K _{q{q^{\prime}  } } $, respectively
 (note that though $\ell_2(t)$ and $\ell'_2(t)$ are  parallel, the map sending one of them to the other fails to do the same with their velocities. 
That is why we cannot use the translation as a  gluing isometry between them).

 Thus,   one can     obtain the desired spacetime $S_f$ by gluing together certain surfaces. In doing this one identifies only points with the sane $t$. Therefore, by  the criterium \eqref{eq:iso}, $S_f$   contains no closed causal curves.




\section{Conclusions}

In summary, we have  demonstrated
 that for any  $d$,  $\alpha <\pi $, and $v<c$
 there is a \emph{causality  respecting}
  spacetime, $S_f$, which describes the  scattering with the impact parameter  $d $ of two strings. Either
   moves in an ``otherwise Minkowski" space with the speed $v$ and has
   the angle deficit $\alpha$.

From the Minkowski space $M$
  \begin{equation*}
\rmd s^2= -\rmd t^2 +  \rmd x^2 + \rmd y^2,\qquad
  t, x, y \in \rea .
\end{equation*}
remove the wedge
\[W\equiv \{ p\in M:\quad t(p)> k|x(p)|\},\quad k=const>1
\]
and glue together the boundaries   \[\partial N_{1,2}
\equiv \{ p\in M:\quad t(p)= k|x(p)|,\ x(p) \lessgtr 0\}.
\]The resulting spacetime, $T_k$, called a \emph{tachyonic string} is similar  to that    considered       in example~\ref{ex:glu}
and describes a superluminal particle.

It is readily seen  that at negative   $t$, $M$ is isometric to $T_k$,  but
the whole spacetimes differ. 
Put another way, the $(t<0)$-region  of a  Minkowski space has infinitely many different (varying  in $k$)  flat extensions.  In fact, it is easy to prove  (for example, by employing the notion of ``loop singularity"
\cite{strstr}) a stronger  fact:\\
 \textbf{Proposition.} \emph{Any}  spacetime $M$, if it has a  flat extension $\tilde M\neq \Cl M$, has infinitely many different  flat extensions each of which contains a (loop) string-like singularity.

Thus, in a theory, where string-like singularities are included,  the uniqueness of evolution of a spacetime is out of the question [the opposite claim  made in  \cite{th} should be taken with  caution: in all appearance the author  implies that of all imaginable  singularities only those considered in examples~\ref{ex:glu},\ref{ex:run} (and their intersections) are allowed   in spacetimes under study].

  So, the existence of $S_f$ is by itself not surprising in the least. Moreover, it is a direct consequence of the theorem proven in \cite{notm}.  What \emph{is} surprising is that $S_f$ turns out to be so simple. In particular, it is orientable and string-like, cf.~\eqref{eq:cov}.


\begin{thebibliography}{20}
\bibitem{strstr}\ssy{S.~Krasnikov}{Unconventional stringlike singularities in flat spacetime}
{Phys.\ Rev.\ D}{76}{2007}{024010}
\bibitem{mee}M.~van de Meent,   \emph{Piecewise flat gravitational waves in 3+1 dimensions}     arXiv:1111.6468 [gr-qc]
\bibitem{SheVi}A. Vilenkin and E. P. S. Shellard \textit{Cosmic strings
and other topological defects}, Cambridge University Press, Cambridge
  \bibitem{Braz}{S.~Krasnikov}{  }{PoS ISFTG},{ \it Quasiregular singularities taken seriously }{2009 }{014},  	 arXiv:0909.4963 [gr-qc].
\bibitem{Gott}\ssy{J. R. Gott III}{Closed timelike curves produced by pairs of moving cosmic
 strings: Exact solutions}{Phys.\ Rev.\ Lett.}{66}{1991}{1126}
\bibitem{Hell}\ssy{C. Hellaby}{The gravitational interaction of conical strings}{GRG}{23}{1991}{767}
\bibitem{djh}\ssy{S. Deser, R. Jackiw, and G. 't Hooft}{Physical cosmic strings do not generate closed timelike curves }{Phys.\ Rev.\ Lett.}{68}{1992}{267}
\bibitem{hg}\ssy{M. P.  Headrick and R. Gott III}{(2+1)-dimensional spacetimes containing closed timelike
curves}{Phys.\ Rev.\ D.}{\bf 50}{1994}{7244}
\bibitem{th}\ssy{G.~'t Hooft}{Causality in (2 + 1)-dimensional gravity}{CQG}{9}{1992}{1335}
 \bibitem{notm}\ssy{S. Krasnikov}{ No time machines in classical general relativity}{Class.\ Quantum Grav.}{19}{2002}{4109 }
  Corrigendum
\emph{ibid.} \textbf{31} (2014) 079503.\\
Krasnikov, Serguei  \emph{Back-in-Time and Faster-than-Light Travel in General
Relativity} (Springer: 2018)
\end{thebibliography}
\end{document}